\documentclass[prl,twocolumn,showkeys,showpacs,superscriptaddress,amsmath,amsfonts]{revtex4-1}
\usepackage{amsfonts}
\usepackage{amssymb}
\usepackage{amsmath}
\usepackage{graphicx}
\usepackage{color}

\usepackage{dcolumn}   % column centering
\usepackage{bm}        % bold math
\usepackage{flafter}

\begin{document}

%\draft

\title{Low-Temperature Phase of the Cd$_2$Re$_2$O$_7$ Superconductor: \\
\textit{Ab initio} Phonon Calculations and Raman Scattering}

\date{June 25, 2019}

\author{Konrad J. Kapcia}
\email[e-mail: ]{konrad.kapcia@ifj.edu.pl}
\affiliation{\mbox{Institute of Nuclear Physics, Polish Academy of Sciences,
ul. W. E. Radzikowskiego 152, PL-31342 Krak\'ow, Poland}}

\author{Maureen Reedyk}
\affiliation{Department of Physics, Brock University, St. Catharines, ON, L2S 3A1, Canada}

\author{Mojtaba Hajialamdari}
\affiliation{Department of Physics, Brock University, St. Catharines, ON, L2S 3A1, Canada}

\author{Andrzej Ptok}
%\email[e-mail: ]{aptok@mmj.pl}
\affiliation{\mbox{Institute of Nuclear Physics, Polish Academy of Sciences,
ul. W. E. Radzikowskiego 152, PL-31342 Krak\'ow, Poland}}

\author{Przemys\l{}aw Piekarz}
%\email[e-mail: ]{piekarz@wolf.ifj.edu.pl}
\affiliation{\mbox{Institute of Nuclear Physics, Polish Academy of Sciences,
ul. W. E. Radzikowskiego 152, PL-31342 Krak\'ow, Poland}}

\author{Fereidoon S.~Razavi}
\affiliation{Department of Physics, Brock University, St. Catharines, ON, L2S 3A1, Canada}

\author{Andrzej M. Ole\'s}
%\email[e-mail: ]{a.m.oles@fkf.mpg.de}
\affiliation{\mbox{Marian Smoluchowski Institute of Physics, Jagiellonian University,
             Prof. S. \.Lojasiewicza 11, PL-30348 Krak\'ow, Poland}}
\affiliation{Max Planck Institute for Solid State Research,
             Heisenbergstrasse 1, D-70569 Stuttgart, Germany}

\author{Reinhard K. Kremer}
\affiliation{Max Planck Institute for Solid State Research,
             Heisenbergstrasse 1, D-70569 Stuttgart, Germany}
%\affiliation{Max-Planck-Institut f\"{u}r Festk\"{o}rperforschung,
%Heisenbergstra$\rm\beta$e 1, D-70569 Stuttgart, Germany}

\begin{abstract}
Using an {\it ab initio} approach, we report a phonon soft mode in the
tetragonal structure described by the space group $I4_{1}22$ of the $1$ K
$5d$ superconductor Cd$_2$Re$_2$O$_7$. It induces an orthorhombic
distortion to a crystal structure described by the space group $F222$
which hosts the superconducting state. This new
phase has a lower total energy than the other known crystal structures
of Cd$_2$Re$_2$O$_7$. Comprehensive temperature dependent Raman
scattering experiments on isotope enriched samples,
$^{116}$Cd$_2$Re$_2{^{18}}$O$_7$, not only confirm the already known
structural phase transitions but also allow us to identify a new
characteristic temperature regime around $\sim 80$ K, below which the
Raman spectra undergo remarkable changes with the development of several
sharp modes and mode splitting. Together with the results of the
\textit{ab initio} phonon calculations we take these observations as
strong evidence for another phase transition to a novel low-temperature
crystal structure of Cd$_2$Re$_2$O$_7$.
\end{abstract}
\smallskip

\maketitle

Superconductivity in materials with non-centro\-symmetric crystal
structures currently attracts broad attention \cite{Smi17,Oku07}.
When parity is not a good quantum number anymore, antisymmetric
spin-orbit coupling may mix spin-singlet with spin-triplet components
within the superconducting (SC) state \cite{Yan08}. Materials
containing heavy $4d$ ($5d$) transition metal elements with large
spin-orbit coupling are especially promising candidates to search for
such states \cite{Zenji}. Strong spin-orbit coupling in correlated
electron system can realize unconventional quantum ground state,
as pointed out recently \cite{Krempa2014,Schaffer2016}.

\begin{figure}[t!]
\includegraphics[width=8.2cm]{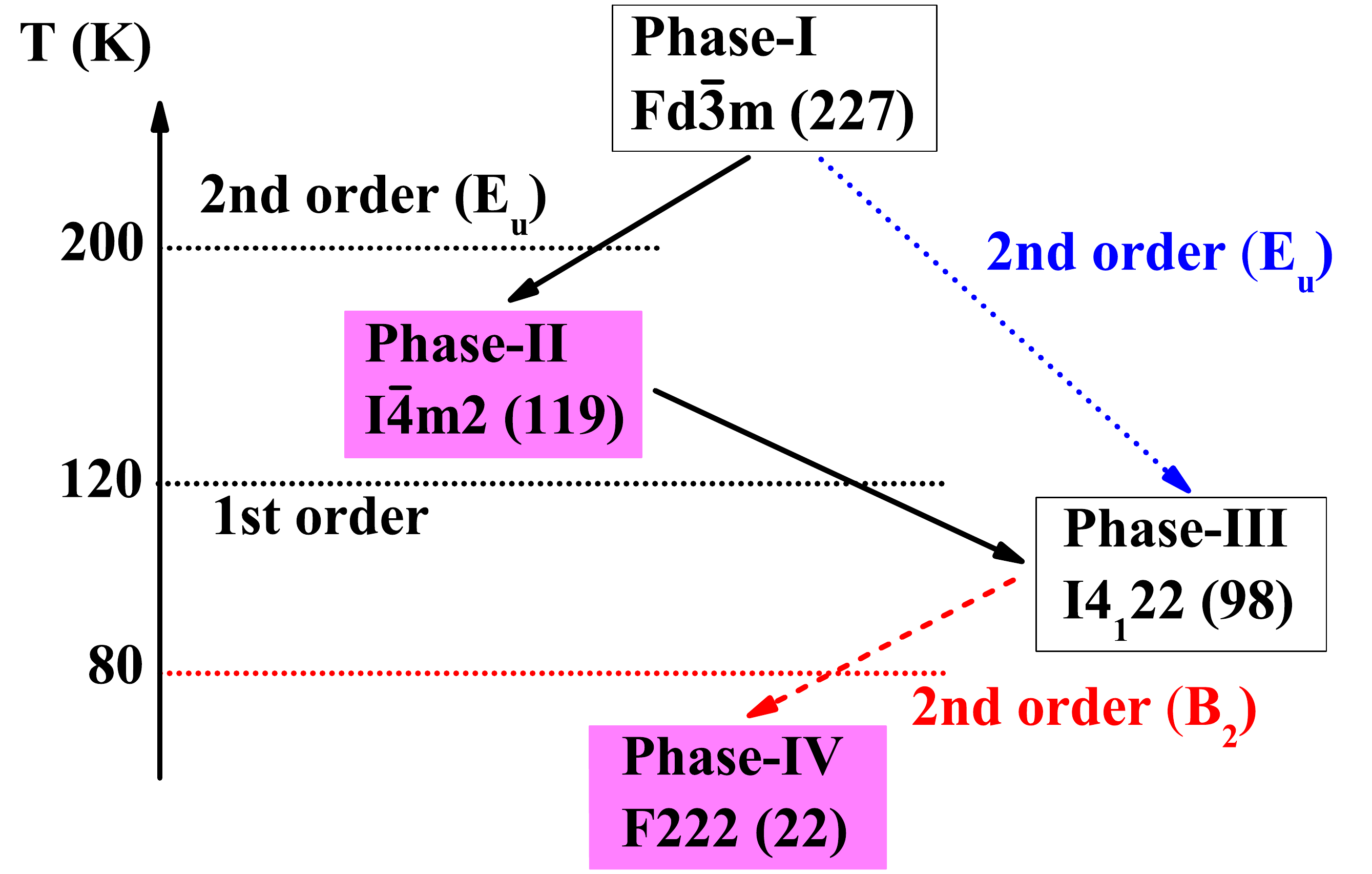}
\caption{\label{fig:1}
Different phases of Cd$_2$Re$_2$O$_7$ (see text).
Solid lines denote the SPTs observed experimentally.
For the second-order SPTs, the irreducible representations of the
associated soft modes are indicated. Magenta boxes denote dynamically
stable phases in our \textit{ab initio} phonon studies.}
\end{figure}

The ternary oxide Cd$_2$Re$_2$O$_7$ (CRO) containing the $5d$
transition metal Re is the first discovered pyrochlore oxide
superconductor with $T_c\simeq 1$ K~\cite{Han01,Sakai2001,Jin2001}.
Among a number of unusual properties reported for CRO, a series of
structural phase transitions (SPTs) is of particular interest since
they essentially determine the low-temperature (LT) structural and
electronic properties \cite{Hir18}. In order to understand the
character of the SC state it proved of especial importance to
characterize the crystal structure at lowest temperatures. At room
temperature (RT) CRO crystallizes in a cubic structure (phase I,
$Fd\bar{3}m$, no. $227$). The 1$^{st}$ SPT to a non-centrosymmetric
tetragonal structure (phase II, $I\bar{4}m2$, no. $119$) takes place
at $\sim 200$~K and a subsequent first order SPT at $\sim 120$~K
to another non-centrosymmetric tetragonal structure (phase III,
$I4_122$, no. $98$) \cite{Yam02,Hir02}, see Fig. \ref{fig:1}.
So far, it is a common belief that this latter phase
is the crystal structure in which superconductivity in CRO occurs.

Both currently known LT tetragonal space groups (phase II and phase
III) are subgroups of the RT cubic space group, see Fig.~\ref{fig:1}.
The structural order parameter transforms according to the irreducible
representation $E_u$ \cite{Ser03,Ser04}. The purpose of this
Letter is to highlight the discovery of a novel non-centrosymmetric
orthorhombic LT phase (phase IV, space group $F222$, no. $22$) for
which we find strong evidence both from density functional theory (DFT)
calculations of the phonon dispersions and detailed temperature dependent
Raman scattering experiments on isotopically enriched single crystals.

Whereas the SPT at $\sim 200$ K leads to pronounced
anomalies, e.g. in the specific heat, the electrical resistivity, the
Pauli susceptibility, or the thermal expansion, the SPT at $\sim 120$ K
is barely perceivable in the bulk properties \cite{Hiroi2002,Tac10}.
A hysteresis in the electrical resistivity points to a 1$^{st}$ order
SPT \cite{Hiroi2002,Hir18}. The impact of the SPT on the electronic
band structure of CRO has been studied within DFT
\cite{Harima2002,Singh2002,Huang2009}. A~semimetallic band structure
with heavy bands near $E_{\rm F}$, heavy hole bands near the zone
boundary, and relatively light electron pockets around the $\Gamma$
point were found. The states near $E_{\rm F}$ are sensitive to spin-orbit
coupling whereas on-site Coulomb interaction $U$ has
negligible effect on Re $5d$ states which are predominantly itinerant.
The SPTs markedly change the carrier density near $E_F$,
as observed in an inversion of the Hall coefficient and rapid decrease
of the spin susceptibility below $\sim 200$ K \cite{Huang2009,Vya02}.
In view of the low carrier density and the heavy band masses,
an excitonic instability of the Fermi surface was suggested as the
origin of the \mbox{I $\rightarrow$ II SPT \cite{Singh2002}}.

The structural distortions induced by the SPTs of CRO are particularly
very minute and difficult to detect by x-ray or neutron scattering
techniques \cite{Hir18}. Raman scattering experiments played
therefore a central role to ascertain the SPTs in CRO, and were also
employed to elucidate the lattice dynamical properties and the role of
electron-phonon coupling \cite{Ken05,Knee2005,Bae06,Raz18,Mat18}.
Below the SPT at $\sim 200$ K, Kendziora \textit{et al.} \cite{Ken05}
observed a peak at zero frequency in the Raman spectra with divergent
intensity which they ascribed to a Goldstone-mode type excitation that
develops due to the breaking of the continuous symmetry at the SPT.
This interpretation has been questioned by Harter \textit{et al.}
\cite{Har18} who argued that the apparent decrease in the Raman center
frequency as the temperature of the SPT is approached from below, could
be due to a reduction of the phonon lifetime,
a conclusion supported by the time-resolved optical reflectivity.
It was proposed that multipolar nematic order rather than a structural
distortion drives the inversion symmetry breaking \cite{Har17}.

Clearly there is a need for further investigation, but regardless of
softening and hardening of phonons and the appearance of new Raman
modes below $200$ K and $120$~K supports symmetry reduction due to
SPTs. The Raman spectra indicated the structural changes at $200$ K and
$120$~K to occur in the Re--O(1) network, with O(1) being the oxygen
atoms at the apices of the ReO$_6$ octahedra \cite{Bae06}. Phonon
spectra have also been studied by IR and ultrafast coherent phonon
spectroscopy \cite{Wang2002,Haj12,Har18}. Optical conductivity studies
revealed that the LT CRO is quite different from a simple metal
\cite{Wang2002} and exhibits anomalous Fermi liquid behavior at LT
\cite{Haj16}. Evidence for strong electron-phonon coupling and enhanced
quasiparticle damping possibly related to a SPT within the SC region
was also found from an analysis of the temperature dependence of the
optical conductivity \cite{Haj12}, and from Andreev
reflection spectroscopy~\cite{Raz15}.

\begin{figure}[t!]
\includegraphics[width=8.2cm]{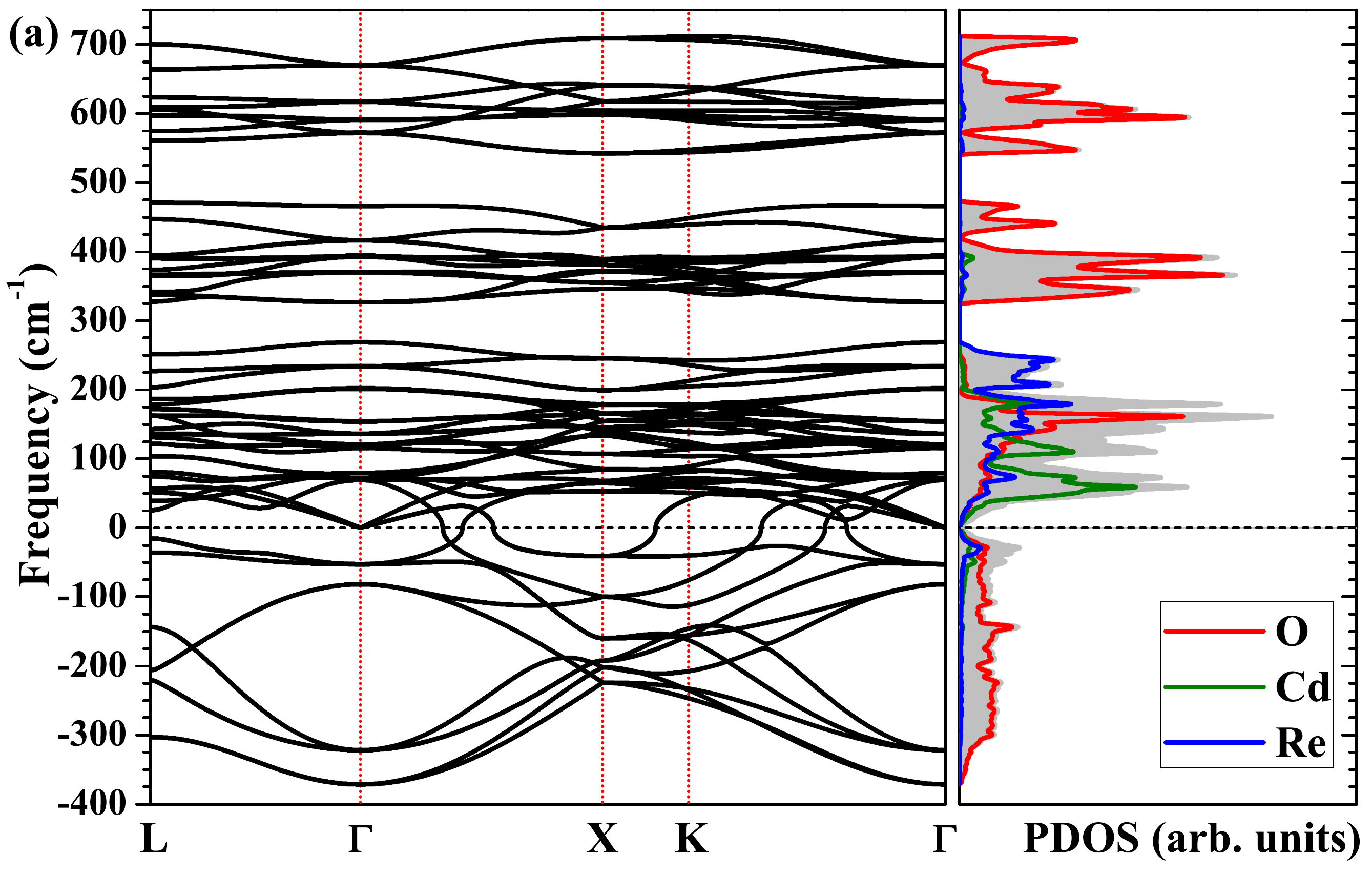}
\vskip -.1cm
\includegraphics[width=8.2cm]{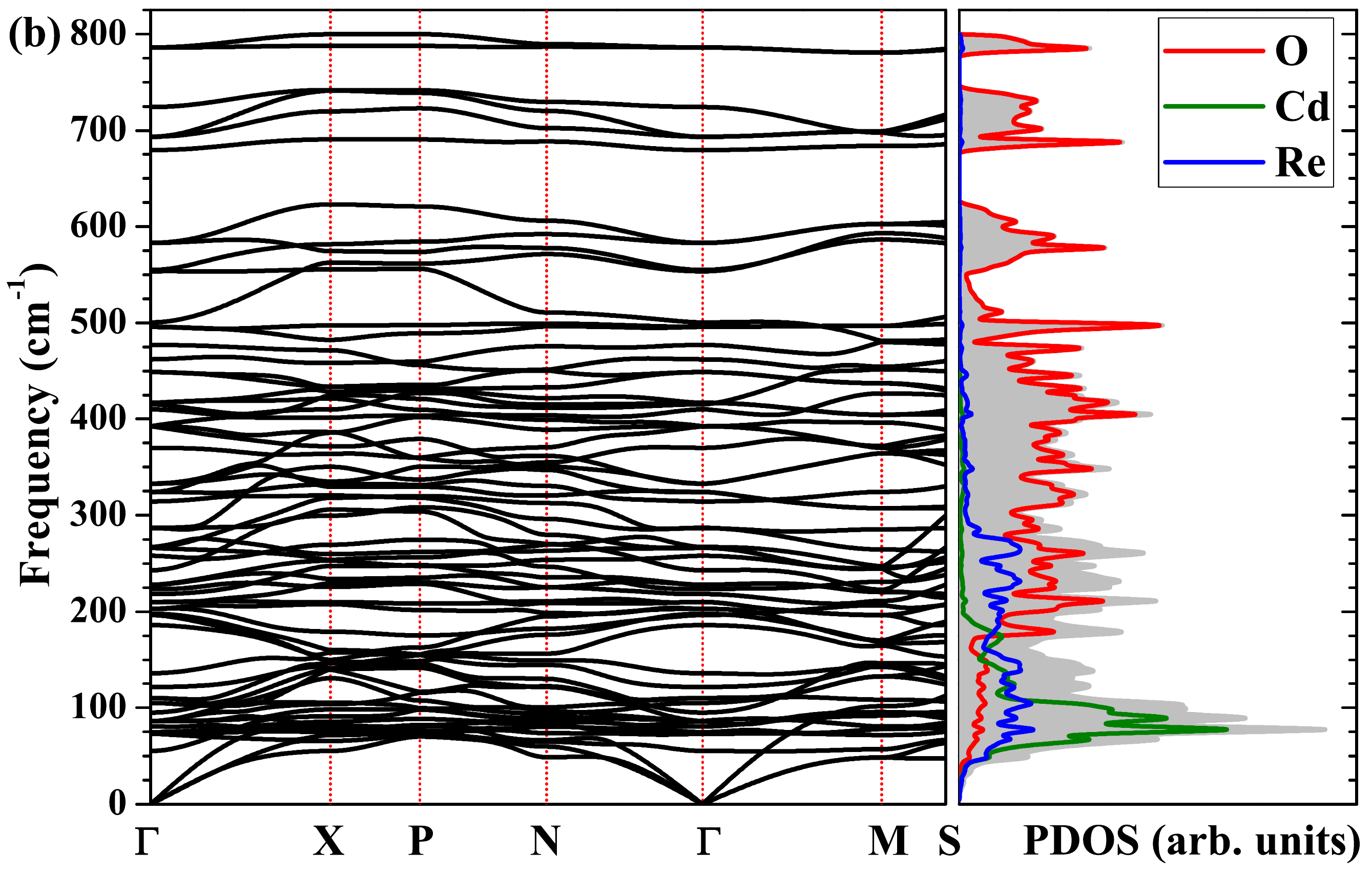}
\vskip -.1cm
\includegraphics[width=8.2cm]{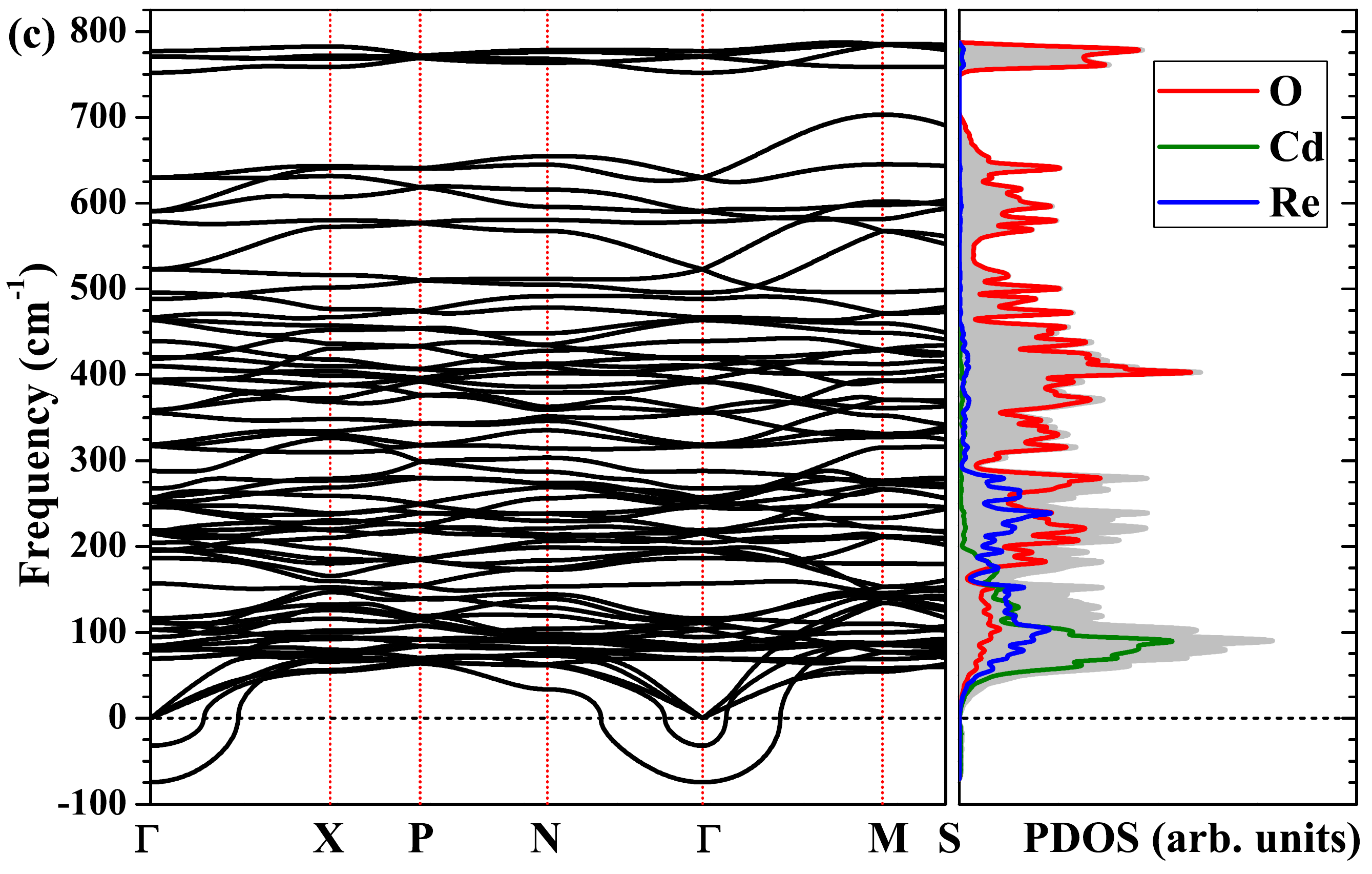}
\vskip -.1cm
\includegraphics[width=8.2cm]{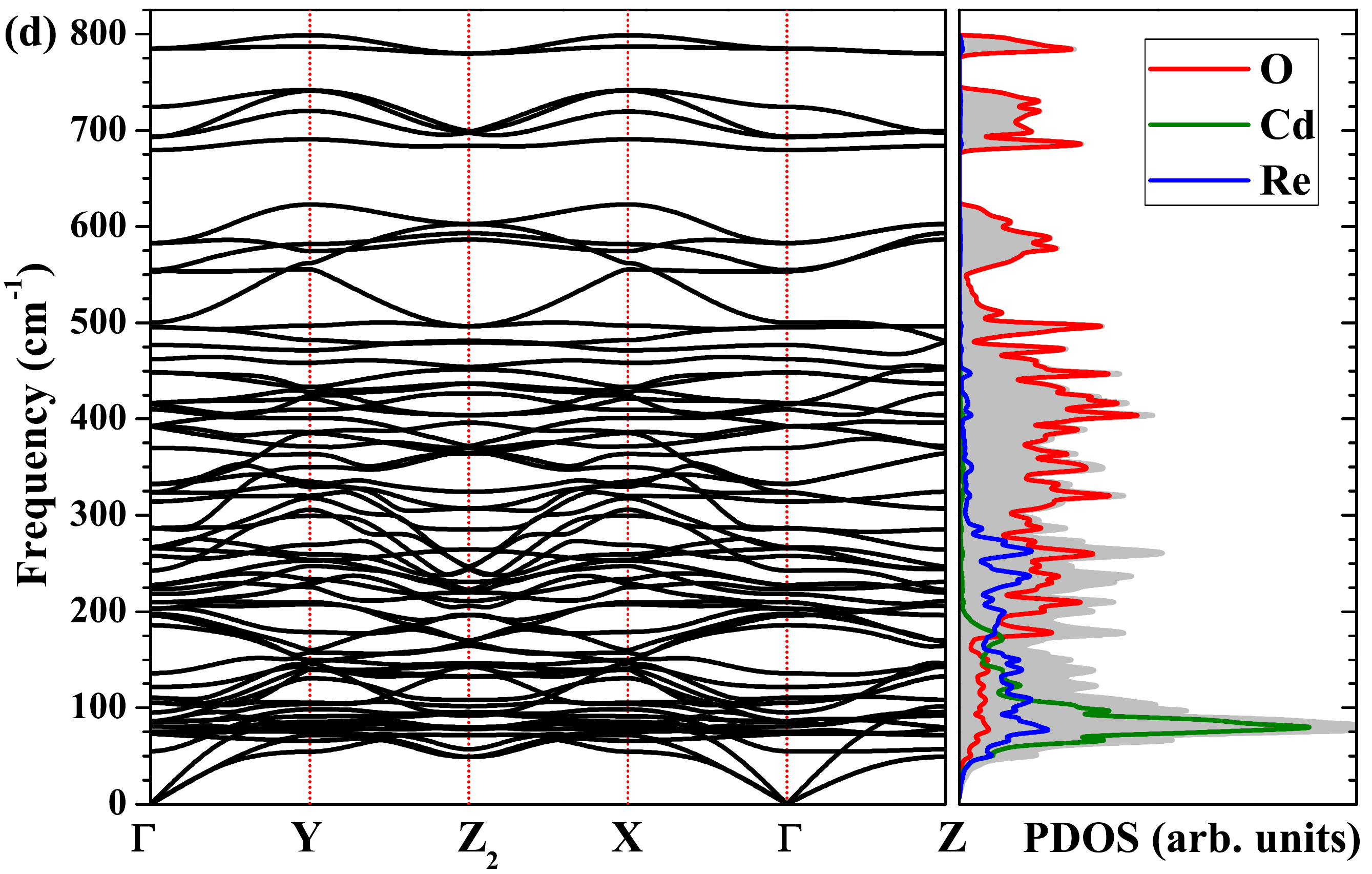}
\caption{\label{fig:2}
The phonon dispersion curves (left) and
partial PDOSs (right) for each phase:
(a) $Fd\bar{3}m$ phase I,
(b) $I\bar{4}m2$ phase II,
(c) $I4_122$ phase III, and
(d) $F222$ phase IV.
\mbox{The gray shaded areas represent the total PDOS.}}
\end{figure}

Our DFT calculations were performed using the projector augmented-wave
method \cite{PAW1,PAW2}, within the generalized gradient approximation
(GGA) \cite{PBE}, implemented in the {\sc Vasp} program package \cite{VASP1}.
A {\bf k}-mesh of 4$\times$4$\times$4 points in the Monkhorst-Pack
scheme was used for integration in the reciprocal space and an energy
cut-off for the plane wave expansion of $500$ eV was applied.
The lattice constants as well as the atom positions were
optimized within the supercells of $88$ atoms, employing the conjugate
gradient technique and energy convergence criteria of $10^{-8}$
($10^{-6}$) eV for electronic (ionic) iterations.

The phonon dispersions and phonon density of states (PDOS) for the
possible structures of CRO were calculated using the direct method
\cite{Par97} implemented in the {\sc Phonon} software \cite{Phonon},
see Fig.~\ref{fig:2}. Since the number of atoms in the primitive cells
of all structures is $22$, all phases have $66$ phonon modes.
The phonon spectra in phase I and phase III exhibit imaginary (soft)
modes indicating that these phases are dynamically unstable at $T=0$.
The most unstable is the cubic structure (phase~I) with imaginary
modes found for each wave vector. The lowest soft phonon is a doubly
degenerate mode with the irreducible representation $E_u$ which agrees
with the group theory analysis \cite{Ser03} and the previous DFT study
\cite{Ser04}. This indicates the displacive character of the
I $\rightarrow$ II SPT connected with the soft mode, although one
cannot rule out a more complex mechanism involving electron-phonon
interactions. Regardless of the mechanism of the SPT, our calculations
show that the total energy of the tetragonal phase II is lowered by
$-2.085$ eV and it is dynamically stable with all phonon branches being
real.

Surprisingly, in the tetragonal phase III
we found the soft modes close to the $\Gamma$ point. The irreducible
representation of the lowest imaginary mode is here $B_2$. Group theory
analysis shows that this mode breaks tetragonal symmetry and generates
the orthorhombic non-centrosymmetric structure (space group $F222$),
called phase IV. We obtained its crystal structure using the
polarization vectors of the $B_2$ soft mode and optimized its crystal
structure parameters, see the Appendix. The
orthorhombic distortion is very weak with the difference between the
$a$ and $b$ lattice parameters about $0.004\%$. The total energy of
this structure is lower than that of the tetragonal phase III by only
$-0.06$ eV and it is very close to phase II ($-0.00005$ eV).
\textcolor{black}{Even though these differences are very small,
modern implementations of the DFT method assure reliability of such
results \cite{Lej16}. To confirm the dynamical stability of phase IV,
we also calculated phonon dispersions and
found only stable (real) modes, see Fig.~\ref{fig:2}(d).}

\begin{figure}[t!]
\begin{center}
\vskip -.5cm
\includegraphics[width=1.25\columnwidth]{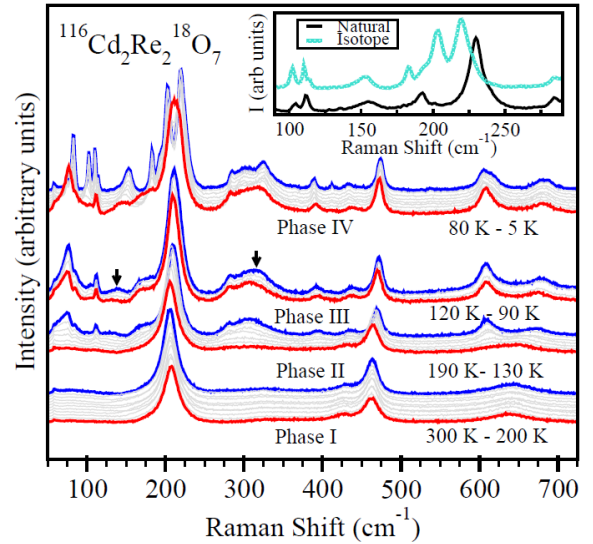}
\end{center}
\vskip -.3cm
\caption{\label{fig:3}
\textcolor{black}{Raman spectra of $^{116}$Cd$_2$Re$_2{^{18}}$O$_7$ in
four characteristic temperature regimes (shifted along the $y$-axis for
clarity). In each regime the red/lower (blue/upper) curve is at the
highest (lowest) temperature, and the gray curves show the evolution in
steps of $10$ K. The arrows mark regions where subtle changes are found
in phase III. The upper inset displays a comparison of two Raman
spectra collected in Phase IV at 25 K of a crystal with the natural
oxygen isotope (99.8\% $^{16}$O) abundance, and of a crystal of
$^{116}$Cd$_2$Re$_2{^{18}}$O$_7$ at 20 K.
}}
\end{figure}

The contribution of particular atoms to the phonon spectra can be
analyzed using the PDOS, see Fig. \ref{fig:2}. Vibrations of the
heavy Cd atoms prevail below $100$ cm$^{-1}$, while Re atoms vibrate
mainly with frequencies $\omega\in(0,300)$ cm$^{-1}$.
Above $300$ cm$^{-1}$, the spectra are dominated by oxygen vibrations
with the cut-off around $700$ ($800$) cm$^{-1}$ in the cubic
(tetragonal) structures. The soft modes found in the cubic phase
largely contribute to the PDOS, with the dominant component of oxygen
vibrations. In contrast, the soft modes have negligible weight in
the total PDOS in the $I4_122$ tetragonal phase.

Oxygen and cadmium isotope-substituted single crystals of CRO have been
prepared by chemical vapor transport \cite{Raz18}. All handling of the
charges was carried out in a water and oxygen free glove box, ensuring
almost complete $^{18}$O substitution in the final product. This
was confirmed by measuring the shift in the frequency of oxygen phonon
modes \cite{Raz18}.
%\textcolor{black}{
Heat capacity measurements on such crystals showed a clear
$\lambda$-type anomaly at the SPT from phase I to phase II at $\sim$204
K and a very faint anomaly at $\sim$114 K indicating the SPT from phase
\mbox{II to phase III}, in good agreement with earlier reports
\cite{Hir18}. The superconducting transition at $T_{c}\sim$ 1 K in 
crystals of $^{116}$Cd$_2$Re$_2{^{18}}$O$_7$ is also seen in the heat 
capacity and ac-susceptibility data (see Figs. 1 and 2 in the Appendix).
Polarized Stokes Raman scattering measurements were
performed in a back-scattering geometry using the $632.8$ nm line of a
Helium-Neon laser at $T\in(5,300)$ K. The measurements were carried
out on the natural growth plane (111) with parallel scattering
geometry. The Raman scattering cross-section contains an $\omega^4$
factor due to Rayleigh scattering which gives a monotonically
decreasing contribution to the background. Its temperature dependence
is subject to the Bose-Einstein factor for thermally excited oscillators.
Therefore, we multiplied our spectra by
\begin{equation}
{1\over{[n(\omega_\sigma)+1]\omega_s^4}}=
{1\over{[n(\omega_\sigma)+1](\omega_o-\omega_\sigma)^4}},
\end{equation}
where $\omega_o$ is the laser frequency, $\omega_\sigma$ is the Raman
shift, $\omega_s$ is the absolute frequency of the scattered light,
and $n(\omega_{\sigma})=1/[\exp(\hbar\omega_{\sigma}/k_{\rm B}T)-1]$ is
the Bose-Einstein factor for thermally excited oscillators.

\begin{figure}[t!]
\includegraphics[width=.99\columnwidth]{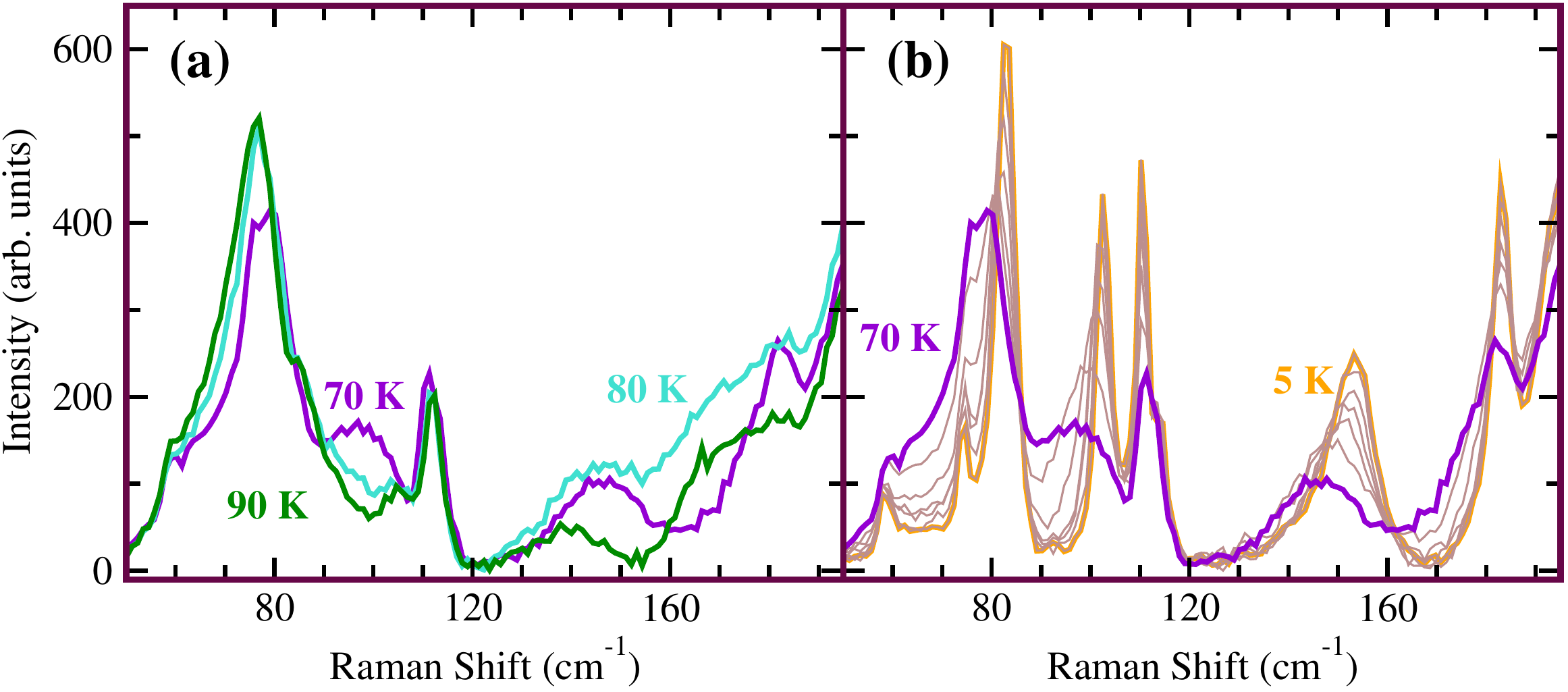}
\caption{\label{fig:4}
Low frequency Raman scattering spectrum of
$^{116}$Cd$_2$Re$_2{^{18}}$O$_7$ showing:
(a) the SPT \mbox{III $\rightarrow$ IV} and
\mbox{(b) the temperature} evolution within phase IV.}
\end{figure}

The Raman spectra of $^{116}$Cd$_2$Re$_2{^{18}}$O$_7$ are displayed for
four characteristic temperature regimes in Fig. \ref{fig:3}.
From $300$ K to $200$ K, i.e., in the cubic $Fd\bar{\rm 3}m$ phase I,
the spectra exhibit five of the six Raman active oxygen modes expected
from a factor group analysis in agreement with previous experiments
\cite{Bae06,Raz18}. The modes near $425$ cm$^{-1}$ and $650$ cm$^{-1}$
have been assigned to $T_{2g}$ modes and the mode near $460$ cm$^{-1}$
to the $A_{1g}$ mode. The mode near $200$ cm$^{-1}$ is comprised of
overlapping $E_g$ and $T_{2g}$ modes~\cite{Mat18}. The absent $T_{2g}$
mode is likely too weak to be resolved, consistent with previous studies
\cite{Knee2005,Bae06,Raz18}. In addition, there is a broad feature
near $350$ cm$^{-1}$, which evolves into a broad mode situated near
$300$ cm$^{-1}$ when the temperature is lowered below $200$ K.

The transition to the tetragonal $I\bar{4}m2$ phase below $200$~K results
in other significant changes to the phonon spectrum including a splitting
of the mode near $640$~cm$^{-1}$, and additional modes appearing below
$175$~cm$^{-1}$. This is consistent with the prediction of a number of
additional Raman active modes according to the factor group analysis.
We observed only very subtle changes at the temperature of the second
SPT from tetragonal $I\bar{4}m2$ to $I4_122$ (II $\rightarrow$ III) near
$115$ K as shown by the arrows in Fig. \ref{fig:3}.

Here we focus on the low frequency Raman scattering in the novel phase
IV. As the temperature is lowered from $80$ K there are further prominent
changes at Raman shifts lower than $250$ cm$^{-1}$ (see Fig. \ref{fig:4}).
Figure \ref{fig:4}(a) shows that the spectrum begins to evolve rapidly
near $\sim$$80$ K from that characteristic for phase III (the $90$ K curve),
while Fig. \ref{fig:4}(b) shows that several sharp peaks develop upon
further lowering of the temperature.

\begin{figure}[b!]
\vskip -.7cm
\begin{center}
\includegraphics[width=1.25\columnwidth]{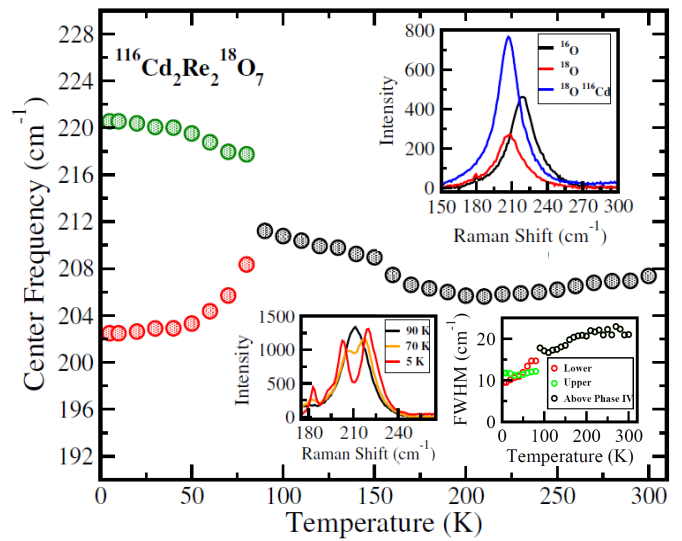}
\end{center}
\vskip -.3cm
\caption{\label{fig:5}
Temperature variation of the center frequency of the $210$ cm$^{-1}$
mode for $^{116}$Cd$_2$Re$_2{^{18}}$O$_7$. Note the splitting at
$\sim80$ K.
%\textcolor{red}{
The lower insets show the mode at temperatures just
above/below $80$~K and at $5$~K and the full-width-half maximum of
the mode above and below the structural phase transition.%}
The upper inset displays this mode measured for a crystal with the
natural oxygen isotope (99.8\% $^{16}$O), for an $^{18}$O isotope
enriched, and for the $^{116}$Cd$_2$Re$_2{^{18}}$O$_7$ at RT.
}
\end{figure}

Further evidence for $T\simeq 80$ K being a characteristic temperature
for a SPT from phase III to phase IV, comes from the behavior of the
$E$ mode near $210$ cm$^{-1}$ in Fig.~\ref{fig:5}. This mode is derived
from oxygen motion as shown by the shift with $^{16}$O isotopic
substitution by $^{18}$O (upper inset in Fig. \ref{fig:5}).
At temperatures above the cubic to
tetragonal SPT at $200$~K, one observes a slight softening of the center
frequency of the mode with decreasing temperature. Below the SPT at
200 K the mode begins to harden as the temperature is lowered, followed
by a remarkable splitting with an onset near $80$ K which reaches
$\sim18$ cm$^{-1}$ at the lowest temperatures, highlighted in the
lower inset in Fig. \ref{fig:5}. As obtained by group theory, the
doubly degenerate Raman mode $E$ in the $I4_122$ structure splits into
two modes ($B_2$ and $B_3$) in the $F222$ structure.
%\textcolor{black}{
Whereas the lower branch fits nicely to a critical power law indicating
a continuous SPT as predicted by the DFT results, the transition to the
upper branch is steeper (see the Appendix for more details).
%\cite{suppl}).}
%\textcolor{red}{
A full splitting of the $210$ cm$^{-1}$ mode has
not been reported before. However, Bae \textit{et al.} \cite{Bae06} and
Knee \textit{et al.} \cite{Knee2005} saw a shoulder and weak modes at
the low-energy side of the $210$ cm$^{-1}$ mode.%}
We consider the distinct splitting of the $210$ cm$^{-1}$
mode as strong support for an additional LT SPT consistent with our
conclusions from the DFT calculations. The symmetry reduction of the
SPT, \mbox{III $\rightarrow$ IV}, leads to eight different oxygen sites
in a unit cell slightly larger than that of the cubic structure, see
Table~I in the Appendix.

\textcolor{black}{It remains puzzling why, so far, we could observe clear
evidence for the SPT from phase III to phase IV only in oxygen isotope
substituted samples, possibly indicating that the SPT at $\sim$80 K can
only be induced or stabilized by increasing the  mass of the oxygen
atoms. A similar observation has been made for SrTiO$_3$ where in
crystal containing only $^{16}$O the transition to a ferroelectric
phase is suppressed by quantum fluctuations, whereas $^{18}$O
substitution induces ferroelectric phases with $T_{\rm c}$'s up to
$\sim$25~K \cite{Ito99,Bus00,Yam04,Row14,Edg15}. Soft-mode behavior of
the vibrational properties has indeed been suggested by preceding Raman
spectroscopy measurements which proposed a Goldstone mode at low
frequencies \cite{Ken05}. We remark that as the temperature is raised
from 5~K to near 80~K, which is well below the SPT from phase II to
phase III at 120~K, that the intensity of the mode investigated by
Kendziora \textit{et al.} grows towards a maximum.}

Summarizing, we have presented remarkable agreement between the
{\it ab initio} calculations of phonon dispersions and the experimental
observations which indicates the existence of a novel orthorhombic
non-centrosymmetric phase IV in Cd$_2$Re$_2$O$_7$ at $T<80$ K.
We believe that our findings will shed new light on the understanding
of the peculiar metallic properties and the occurrence of
the superconducting phase in Cd$_2$Re$_2$O$_7$.

\textit{Acknowledgments.}--- The authors are grateful to
Krzysztof Parlinski for the very insightful discussions and comments.
This work was supported by the National Science Centre (NCN, Poland)
under grants 2017/24/C/ST3/00276 (K.J.K.), 2017/25/B/ST3/02586 (P.P.),
and 2016/23/B/ST3/00839 (A.M.O.)
and the Natural Sciences and Engineering Research Council (NSERC)
of Canada grants DDG-2017-00043 (M.R) and RGPIN-2018-04438 (F.S.R).
A.~M.~Ole\'s is grateful for the Alexander von Humboldt
Foundation Fellowship \mbox{(Humboldt-Forschungspreis).}

\appendix

\section{APPENDIX:\\
Low-Temperature Phase: % of Cd$_2$Re$_2$O$_7$ Superconductor: \\
\textit{Ab initio} Phonon Calculations and Raman Scattering }

%\date{\today}

%\begin{abstract}
In Section I of this Supplemental Material, we present some additional
information on the structural and on the superconducting phase
transitions in $^{116}$Cd$_2$Re$_2{^{18}}$O$_7$ obtained from experiment.
In Section II we present the structural data of Cd$_2$Re$_2$O$_7$
superconductor as obtained from experiment [1] and from the density
functional theory calculations.
%\end{abstract}
%\maketitle

\begin{figure}[t!]
\includegraphics[width=\columnwidth]{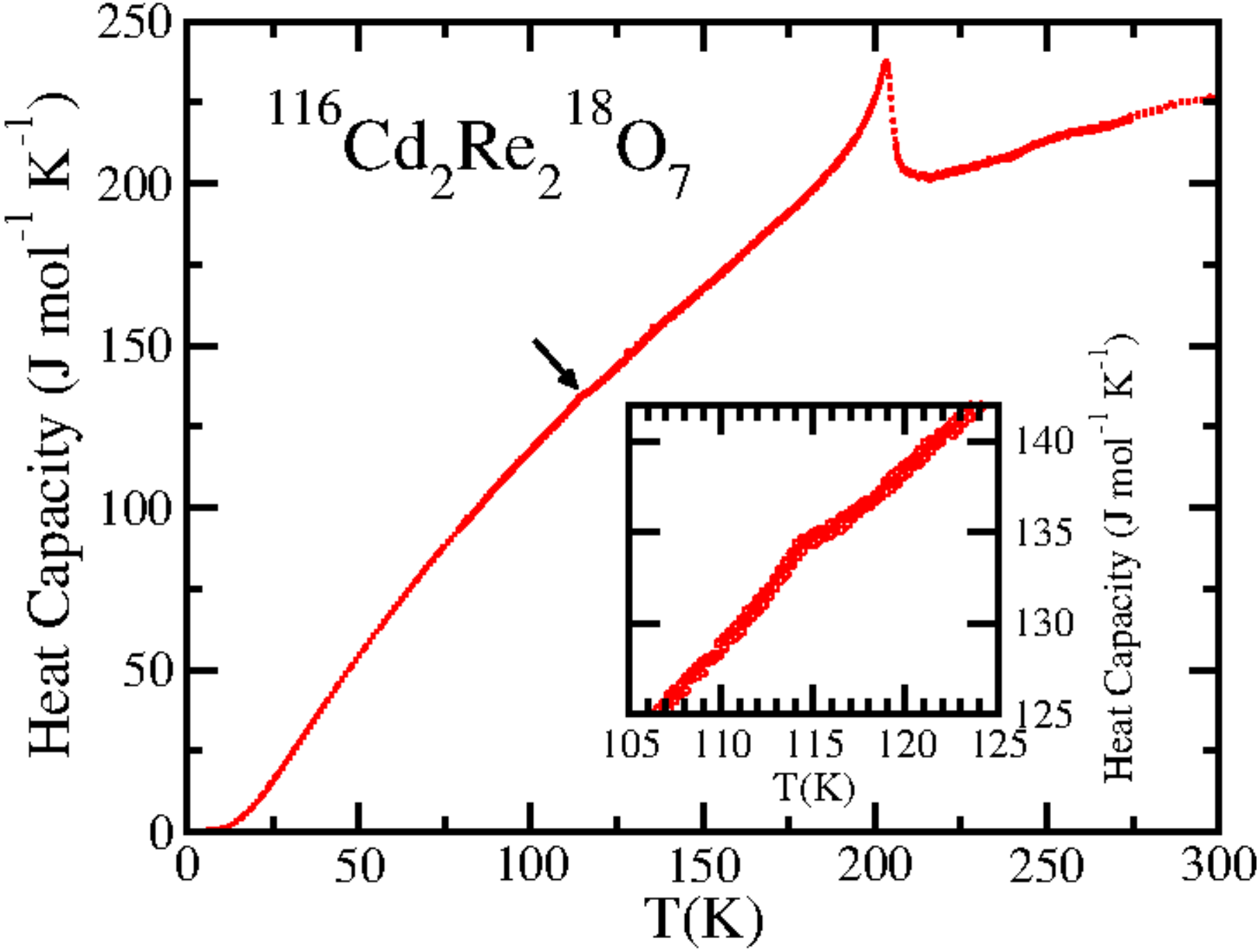}
\caption{\label{HeatCap}
Specific heat capacity of a crystal of the isotope enriched material,
$^{116}$Cd$_2$Re$_2{^{18}}$O$_7$. Anomalies at $\sim$204 K and at
$\sim$114 K (arrow and inset) prove the SPT from phase I to phase II
and from phase II to phase III, respectively.}
\end{figure}

\begin{figure}[b!]
\hskip -.7cm
\includegraphics[width=1.2\columnwidth]{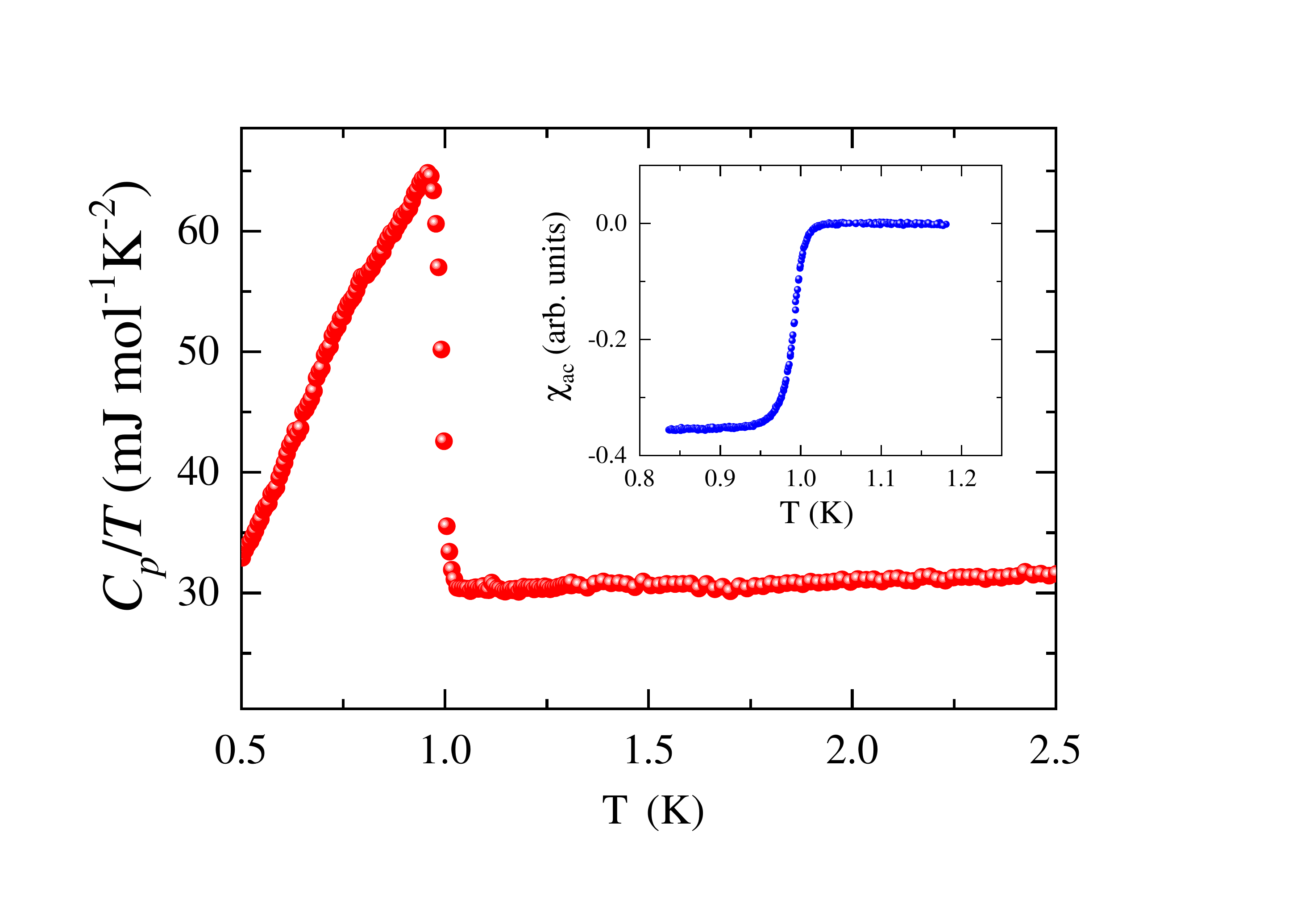}
\vskip -.7cm
\caption{\label{cvsup}
Specific heat capacity, $C_p/T$, of a crystal of the isotope enriched
material, $^{116}$Cd$_2$Re$_2{^{18}}$O$_7$, at low temperature.
A characteristic $\lambda$-type anomaly at the onset of the
superconducting phase at $T_c\sim$1 K is observed. The inset shows the
\mbox{ac-susceptibity} with a distinct drop at $T_{\rm c}$.}
\end{figure}

\subsection{A. Experimental details}

In this Section we present detailed results of the characterization of
the isotope substituted crystals, $^{116}$Cd$_2$Re$_2{^{18}}$O$_7$,
used in the Raman investigation with emphasis on the structural phase
transitions (SPTs).
Figure \ref{HeatCap} displays the heat capacity measured on a crystal
of $^{116}$Cd$_2$Re$_2{^{18}}$O$_7$. The well defined $\lambda$-type
anomaly at the SPT from phase I to phase II is seen at $\sim$204 K and
a very faint anomaly at $\sim$114 K indicating the SPT from phase II to
III is revealed, in good agreement with earlier reports \cite{Hir18}.

The heat capacity measurements also revealed the anomaly at the
$T_{\rm c} \sim$1 K shown in Fig. \ref{cvsup}.

\begin{figure}[t!]
%\hskip -.9cm
\includegraphics[width=\columnwidth]{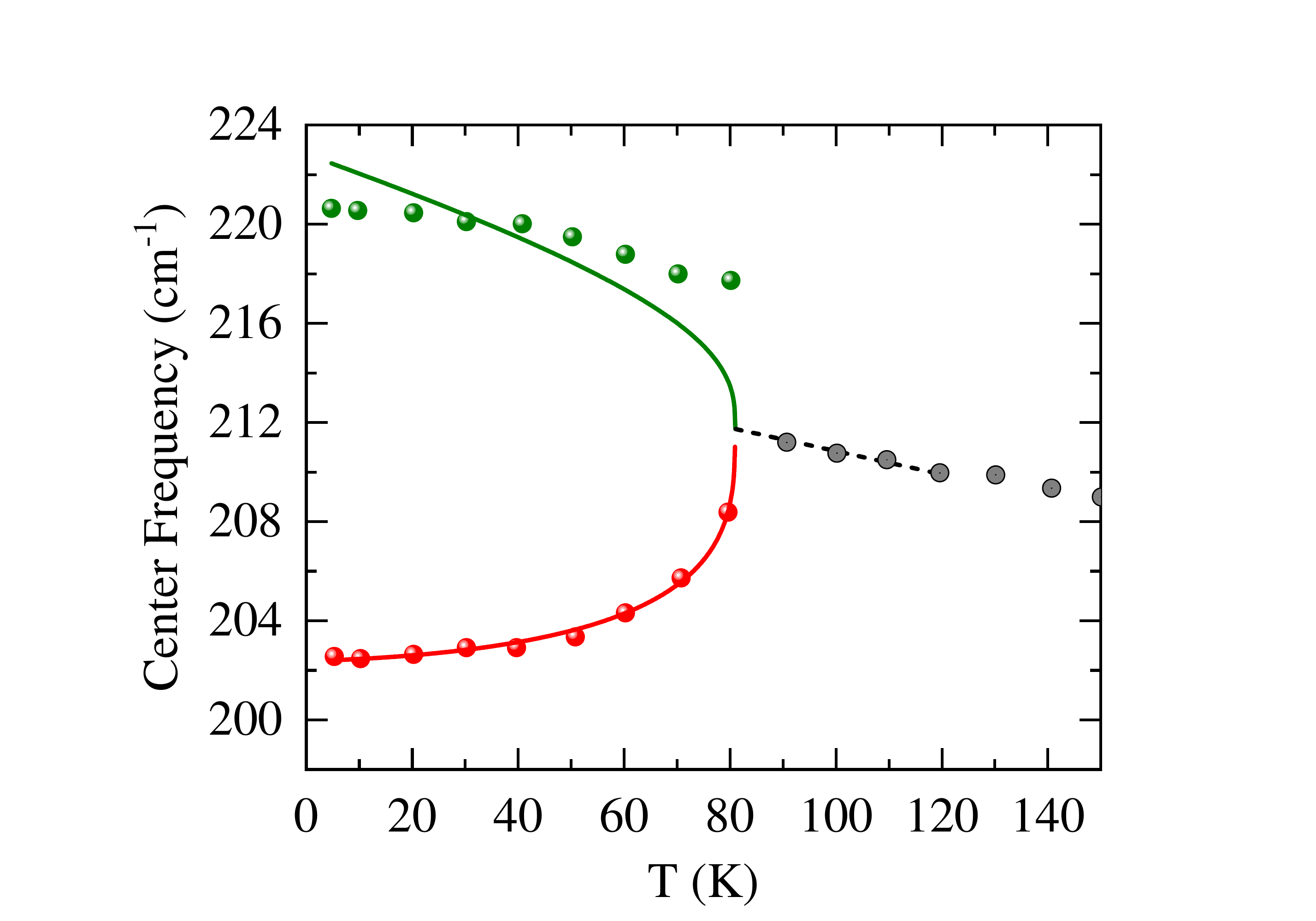}
\caption{\label{CritPow}
Splitting of the $E$ mode observed near 210 cm$^{-1}$ versus
temperature for $^{116}$Cd$_2$Re$_2{^{18}}$O$_7$. The solid lines
represent fits of the center frequencies with a critical power law
(see text) indicating a transition temperature of $\sim$81 K.}
\end{figure}

Figure \ref{CritPow} shows the splitting of the $E$ mode observed near
210 cm$^{-1}$ below $\sim$80 K. The red (lower) and green (upper)
solid line represents the results of a fit of the center frequencies
with a critical power, $t^{\beta}$, where
\mbox{$t=(T_{\rm c} -T)/T_{\rm c}$} and $\beta=0.33$ is the critical
exponent. $T_{\rm c}$ was obtained as $\sim$81 K from a fit of the
lower branch and kept identical for the upper branch.

\begin{table*}[t!]
\caption{\label{tab:str} Lattice parameters, atomic positions, and total energies
in the crystal structures of Cd$_2$Re$_2$O$_7$.}
\begin{ruledtabular}
\begin{tabular}{l|lll|lll|}
& \multicolumn{3}{c|}{Experimental data \cite{Hua09}} & \multicolumn{3}{c|}{DFT calculations} \\
\hline
\hline
& \multicolumn{6}{c|}{Phase-I: $Fd\bar{3}m$ (227)}\\
\hline
& \multicolumn{3}{c|}{$a=b=c=10.2261(5)$ \AA} & \multicolumn{3}{c|}{$a=b=c= 10.3816$ \AA} \\
& \multicolumn{3}{c|}{} & \multicolumn{3}{c|}{$V_{I}=1118.90$ \AA$^3$, $E_I=0$} \\
Re & 0.2500 & 0.7500 & 0.5000      & 0.0000 & 0.2500 & 0.2500 \\
Cd & 0.5000 & 0.5000 & 0.5000      & 0.5000 & 0.5000 & 0.5000 \\
O(1) & 0.3152(9) & 0.6250 & 0.6250 & 0.3141 & 0.1250 & 0.1250\\
O(2) & 0.6250 & 0.6250 &  0.6250   & 0.6250 & 0.6250 & 0.6250 \\
\hline
\hline
& \multicolumn{6}{c|}{Phase-II: $I\bar{4}m2$ (119)}\\
\hline
& \multicolumn{3}{c|}{$a=b=7.2312(3)$ \AA, $c=10.2257(4)$ \AA} & \multicolumn{3}{c|}{$a=b=7.3420$ \AA, $c=10.4101$ \AA} \\
& \multicolumn{3}{c|}{} &\multicolumn{3}{c|}{$a'=b'=10.38313$ \AA ($a'=\sqrt{2}a$), $c'=10.41010$ \AA} \\
 & \multicolumn{3}{c|}{} & \multicolumn{3}{c|}{$V'_{II}=1122.31$ \AA$^3$, $E_{II}=-2.08510$ eV} \\
 Re & 0.2471(2) & 0.0000 & 0.87294(15)	 & 0.2510 & 0.0000 & 0.8760 \\
 Cd & 0.0000 & 0.2471(5) & 0.6259(3) 	 & 0.0000 & 0.2372 & 0.6386 \\
 O(1) & 0.3059(15) & 0.1941(15) & 0.7500 & 0.3000 & 0.2000 & 0.7500 \\
 O(2) & 0.0000 & 0.0000 & 0.8026(16) 	 & 0.0000 & 0.0000 & 0.7947 \\
 O(3) & 0.1889(16) & 0.1889(16) & 0.0000 & 0.1795 & 0.1795 & 0.0000\\
 O(4) & 0.5000 & 0.0000 & 0.9317(19) 	 & 0.5000 & 0.0000 & 0.9189 \\
 O(5) & 0.0000 & 0.0000 & 0.5000 		 & 0.0000 & 0.0000 & 0.5000\\
 O(6) & 0.0000 & 0.5000 & 0.7500 		 & 0.0000 & 0.5000 & 0.7500\\
 \hline
  \hline
& \multicolumn{6}{c|}{Phase-III: $I4_{1}22$ (98)}\\
\hline
& \multicolumn{3}{c|}{$a=b=7.2313(3)$ \AA, $c=10.2282(6)$ \AA} & \multicolumn{3}{c|}{$a=b=7.3537$ \AA, $c=10.3729$ \AA} \\
& \multicolumn{3}{c|}{} &
\multicolumn{3}{c|}{$a'=b'=10.3997$ \AA ($a'=\sqrt{2}a$), $c'=10.3729$  \AA} \\
& \multicolumn{3}{c|}{} & \multicolumn{3}{c|}{$V'_{III}=1121.87$ \AA$^3$, $E_{III}=-2.02515$ eV } \\
 Re & 0.2500 & 0.9967(3) & 0.8750 & 	   0.2500 & 0.0020 & 0.8750\\
 Cd & 0.5041(6) & 0.2500 & 0.1250 & 	   0.5229 & 0.2500 & 0.1250\\
 O(1) & 0.1880(20) & 0.1880(20) & 0.0000 & 0.1711 & 0.1711 & 0.0000\\
 O(2) & 0.5000 & 0.0000 & 0.9396(14) & 	   0.5000 & 0.0000 & 0.9424\\
 O(3) & 0.1970(20) & 0.8030(20) & 0.0000 & 0.2032 & 0.7968 & 0.0000\\
 O(4) & 0.5000 & 0.5000 & 0.0000 &		   0.5000 & 0.5000 & 0.0000\\
 \hline
 \hline
& \multicolumn{6}{c|}{Phase-IV: $F222$ (22)}\\
\hline
& \multicolumn{3}{c|}{} & \multicolumn{3}{c|}{$a=10.3832$ \AA, $b=10.3828$ \AA, $c=10.4101$ \AA} \\
& \multicolumn{3}{c|}{} & \multicolumn{3}{c|}{$V_{IV}=1122.28$ \AA$^3$, $E_{IV}=-2.08515$ eV} \\
Re & & & & 0.124500 & 0.124502 & 0.1260 \\
Cd & & & & 0.63141 & 0.63137 & 0.6114 \\
O(1) & & & & 0.5000 & 0.5000 & 0.1689 \\
O(2) & & & & 0.0706 & 0.2500 & 0.2500 \\
O(3) & & & & 0.2500 & 0.4295 & 0.2500 \\
O(4) & & & & 0.5000 & 0.2000 & 0.5000 \\
O(5) & & & & 0.2000 & 0.5000 & 0.5000 \\
O(6) & & & & 0.2500 & 0.2500 & 0.0447 \\
O(7) & & & & 0.5000 & 0.5000 & 0.5000 \\
O(8) & & & & 0.7500 & 0.7500 & 0.7500
\end{tabular}
\end{ruledtabular}
\end{table*}
%\eject 

\subsection{B. Structural data}

In Table I we present the structural data as obtained from
experiment \cite{Hua09} and from the present density functional
theory (DFT) calculations: $a$, $b$, and $c$ denote the lattice
parameters, while $a'$, $b'$, and $c'$ denote the sizes of the
supercell used for the calculations. Here $V$ ($V'$) is the volume of
the unit cell (supercell). We have included also the total energies
given per one supercell with 88 atoms for all studied phases.

\end{document}